\documentclass[epjST]{svjour}
\usepackage{graphics}
\usepackage{xcolor}     
\usepackage{soul}       
\begin{document}
\title{{A new setup for experiments with ultracold Dysprosium atoms}}
\author{E.~Lucioni\inst{1,3} \and G.~Masella\inst{1} \and A.~Fregosi\inst{2} \and C.~Gabbanini\inst{1} \and S.~Gozzini\inst{1} \and A.~Fioretti\inst{1}\fnmsep\thanks{\email{andrea.fioretti@ino.it}} \and L.~Del~Bino\inst{3} \and J.~Catani\inst{3,4} \and G.~Modugno\inst{3,4} \and M.~Inguscio\inst{3}}
\institute{Istituto Nazionale di Ottica, CNR., S.S.  ``{\it A.~Gozzini} ''
 di Pisa, via Moruzzi 1, 56124 Pisa, Italy
\and Dipartimento di Fisica, Universit$\grave{\rm a}$ di Pisa, Largo B. Pontecorvo, 56127 Pisa, Italy
\and LENS and Dip. di Fisica e Astronomia, Universit$\grave{\rm a}$ di Firenze, 50019 Sesto Fiorentino, Italy
\and Istituto Nazionale di Ottica, CNR, S.S. Sesto Fiorentino, 50019 Sesto Fiorentino, Italy}
\abstract{
In the domain of  quantum degenerate atomic gases, much interest has been raised recently by the use of Lanthanide atoms with large magnetic moments, in particular Dysprosium and Erbium. These species have been successfully {brought} to quantum degeneracy and are now excellent candidates for quantum simulations of {physical phenomena due to} long-range interactions. In this short article, we report on the  progresses in the construction of {a new} experiment {on} Bose-Einstein condensation of Dysprosium atoms. After completing the vacuum and the laser setups, a magneto-optical trap on the narrow $626\, {\rm nm}$ $^{162}$Dy transition has been realized and characterized. The prospects {for future experiments} are briefly discussed.
} 
\maketitle
%
{Quantum degenerate} dipolar systems  are nowadays among the most interesting systems in {quantum} physics because they provide, to a very large extent, a clean and controlled experimental environment where {long-range, anisotropic interactions can be finely tuned against short-range, isotropic ones}. {Although the dipolar interaction strength is much lower than that achievable in molecular systems, atomic samples are suitable to study a variety of phenomena thanks to the absence of large inelastic losses and to the immediacy of application of well-developed experimental techniques.  Dipolar interactions in ultracold atoms, according to theoretical predictions~\cite{Lahaye2009,Trefzger2011,Baranov2012}, give rise to a wealth of peculiar quantum phenomena and exotic quantum phases, encompassing super-solids, quasi-crystals, frustrated crystals and self-assembled structures.} These phenomena are due to the combination of the long-ranged and anisotropic nature of such interactions.  Dipolar gases in optical lattices {may} widen the {existing} possibilities {for the} quantum simulation of condensed matter-like physics~\cite{Lewenstein2007,Bloch2008}.

Among open-shell Lanthanides having the largest magnetic interaction, quantum degeneracy has been recently attained for Dysprosium~\cite{Lu2011a,Lu2012a} and Erbium~\cite{Aikawa2012,Aikawa2014}, both nicely providing bosonic and fermionic isotopes in large natural abundance. These systems have shown a very rich collisional dynamics~\cite{Baumann2014,Maier2015a,Maier2015b} due to their large spin and orbital angular momentum and, most of all, spectacular phase transitions~\cite{Kadau2016,Chomaz2016} { obtained by fine Feshbach tuning of the contact interaction with respect to the dipolar one.} 
Finally, a first quantum simulation of the extended Hubbard model has been realized with a dipolar condensate in an optical lattice~\cite{Baier2016}.

The starting point and working horse of all systems leading to quantum degeneracy in dilute gases is a magneto-optical trap (MOT). In open-shell Lanthanide atoms, given the complex electronic structure, it {was} far from being obvious that the MOT scheme would work {but} it has been successfully demonstrated initially with Erbium atoms~\cite{Berglund2008}. The presence of several possible cycling transitions with different linewidths~\cite{Berglund2008,Lu2010,Frisch2012,Maier2014} offers the possibility of tailoring the cooling force and the final temperature in order to optimize the subsequent trapping in optical traps and evaporative cooling to quantum degeneracy.

We report here on the realization of a magneto-otical trap for $^{162}$Dy atoms on the intermediate linewidth transition at $626\, {\rm nm}$. {We briefly describe our experimental setup and the characterization of the loading and trapping processes}. Our results confirm those of a recent extended study of a Dy MOT~\cite{Dreon2016} and represent a nice starting point for evaporative cooling to quantum degeneracy.
%

\begin{figure}[htbp]
\centering{
\resizebox{0.85\columnwidth}{!}{
  \includegraphics{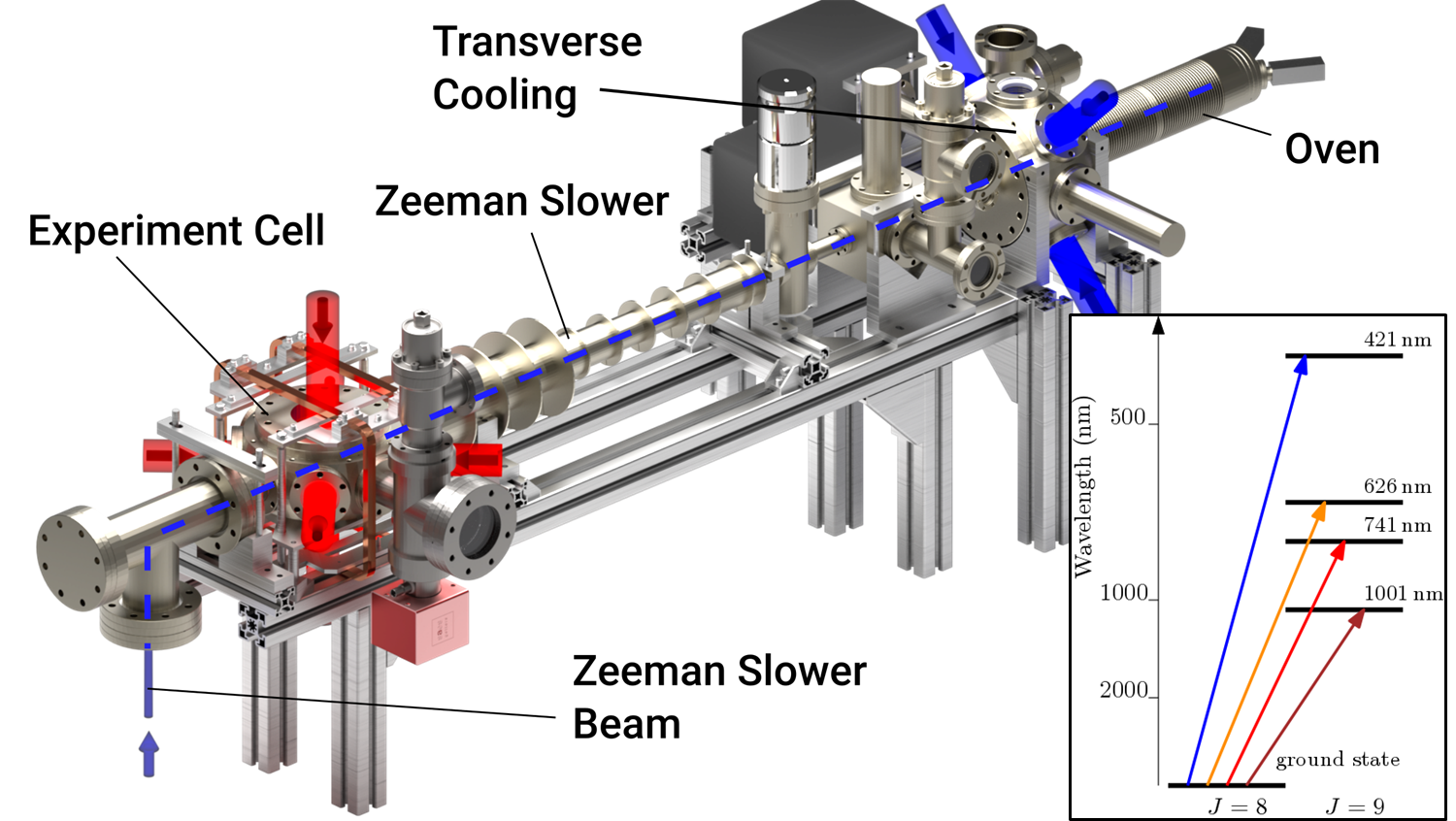} }
}
\caption{Sketch of the experimental apparatus. Laser beams are indicated by blue ($421\, {\rm nm}$) and red ($626\, {\rm nm}$) arrows. In the inset, the $J\rightarrow J+1$ transitions from Dy ground state are shown. The two transitions $ $ at $421\, {\rm nm}$ and $626\, {\rm nm}$ are used for slowing and trapping, respectively.}
\label{fig:apparatus}       
\end{figure}

Disprosium melts at $1407\,^{\circ}$C and therefore a high temperature oven is necessary to obtain a reasonable vapour pressure, thus requiring substantial laser slowing in a Zeeman Slower (ZS).
The experimental apparatus, shown in Fig.~\ref{fig:apparatus} along with Dy relevant electronic transitions, is made up of three main sections. In the first section, {evacuated by a $20\, {\rm l/s}$ ion pump (Varian VacIon),} an atomic beam is sourced from an effusive cell heated up to $1200\,^{\circ}$C. In order to exit the cell, the Dy atoms pass throught a $3\, {\rm cm}$-long collimation tube with a diameter of $4\, {\rm mm}$. In this way forward emission is enhanced by reducing by a factor 10 the atomic beam divergence and a longer lifetime of solid Dysprosium in the crucible is guaranteed. A short stainless steel tube ($4 \, {\rm cm}$ diameter) after the cell is nevertheless used to prevent Dy atoms to stick on the side windows.  The atomic beam is further collimated by a transverse cooling stage performed on the strongest transition at $421\, {\rm nm}$ (linewidth $\Gamma_{421}= 2 \pi \times 32.2\, {\rm MHz}$). Transverse cooling beams are elliptically shaped ($3.5\, {\rm mm}$ and $8\, {\rm mm}$ vertical and horizontal waists), light is red detuned by 0.3$\Gamma_{421}$ from the atomic transition and its intensity is higher than the saturation intensity ($I_{421}=3 \times I_{\rm sat}^{421}$ where $I_{\rm sat}^{421}=56.5\, {\rm mW/cm}^2$) for a total power of $75\, {\rm mW}$ in the beams. We do not observe a saturation of the transverse cooling effect with the power and we suppose that a higher power would further increase the MOT loading efficiency. 

A differential pumping stage separates the first section of the apparatus from the second one. {This is composed of two narrow tubes (each $5\, {\rm mm}$ diameter and $81\, {\rm mm}$ length) pumped in between by another $20\, {\rm l/s}$ ion pump (Varian VacIon). Together with the ZS tube, they allow a differential vacuum between the transverse cooling cell and the science cell larger than three orders or magnitude}
In this section the fraction of the atomic beam with velocity lower than about $550\, {\rm m/s}$ is slowed down  to a final velocity of a few meters per second in a $52\, {\rm cm}$ long spin-flip Zeeman slower operating on the strong $421\, {\rm nm}$ transition as well. The ZS beam has a power of $150\, {\rm mW}$, is red detuned by 27$\Gamma_{421}$ ($2\pi \times1.05\, {\rm GHz}$) from the atomic transition, {enters the ZS through an aluminum mirror positioned inside the vacuum apparatus at 45$^{\circ}$~\cite{Aikawa2012} in order to avoid Dy coating on the window,} and is focused of the effusive cell aperture to better match the atomic beam shape.

The third section consists of a stainless steel octagonal chamber where atoms from the slowed beam are {captured} in a magneto-optical trap (MOT) operating on the narrower atomic transition at $626\, {\rm nm}$ with a linewidth $\Gamma_{626}= 2 \pi \times 136\, {\rm kHz}$. The  trapping beams have large waists (1.2-$1.8\, {\rm cm}$) and peak intensities of about $I_{626}=120 \times I_{\rm sat}^{626}$ where $I_{\rm sat}^{626}=72\, \mu$W/cm$^2$) for a total power of $140\, {\rm mW}$ in the six beams, and a detuning $\Delta$ in the 1-$6\, {\rm MHz}$ range. The magnetic field gradient is about $1.4\, {\rm G/cm}$ along the strong axis.
{This final section is evacuated by a NEXTorr D 100-5 hybrid pump ($100\, {\rm l/s}$ getter + $6\, {\rm l/s}$ ion pump). This choice has the advantage of providing large pumping speed with reduced magnetic fields near the atoms.}
\begin{figure}[htbp]
\centering{
\resizebox{0.5\columnwidth}{!}{
  \includegraphics{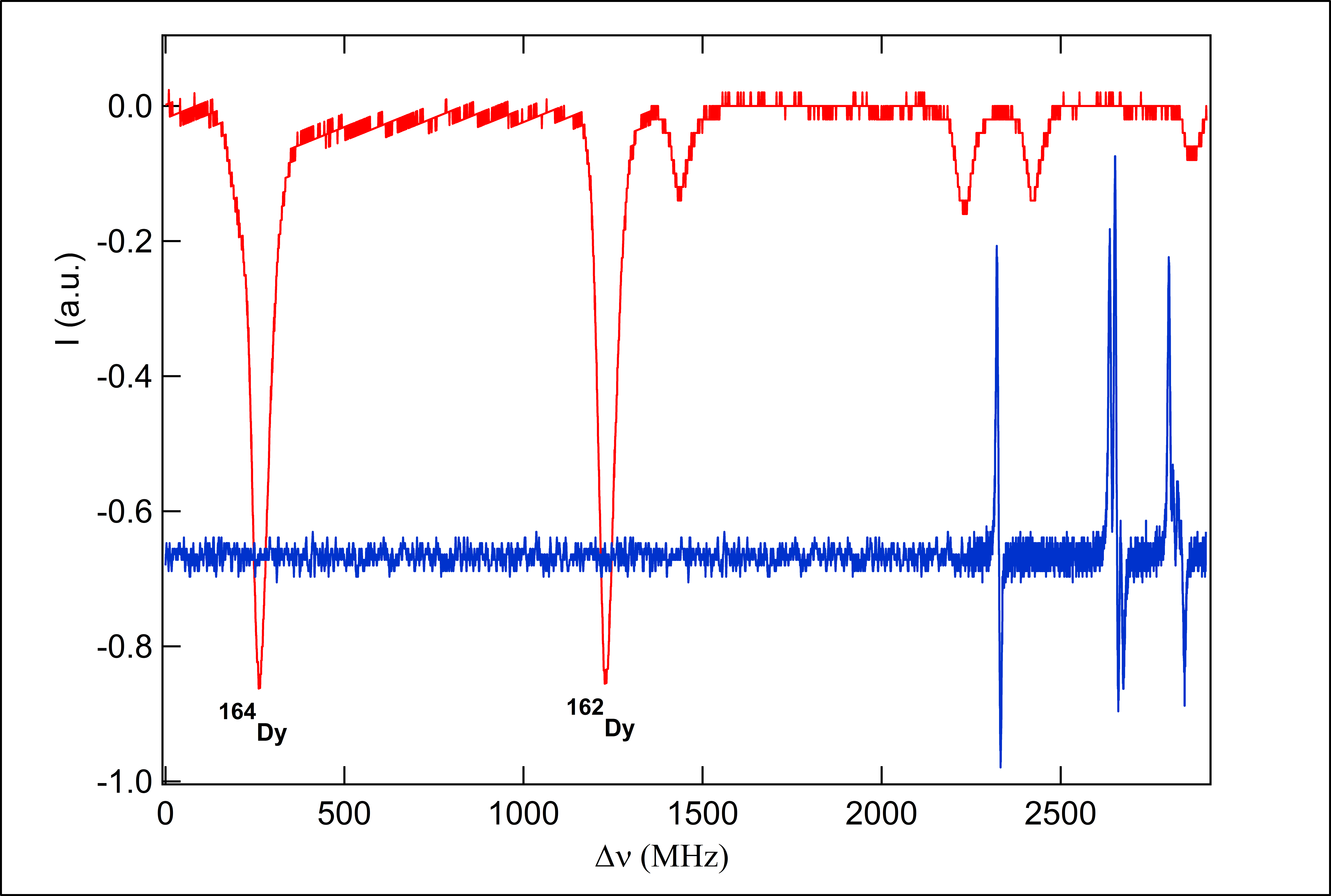} }
}
\caption{Fluorescence spectrum (upper red curve) of the Dy atomic beam and error signal (lower blue curve) of the saturated absorption of Iodine as a function of the laser frequency near $626\, {\rm nm}$. The largest peaks from the left come from the two bosonic species $^{164}$Dy and $^{162}$Dy, while the smaller ones belong to the fermions. We lock the laser to the left-most line of Iodine in the figure.}
\label{Fig:lock}       
\end{figure}

In order to generate light at $421\, {\rm nm}$ we frequency-double the light emitted from a Ti:Sa laser (SolsTiS by M Squared, 3W output at $800\, {\rm nm}$) in a bow-tie cavity containing a LBO crystal ($3\times 3 \times 15\, {\rm mm}^3$, $27.8\,^{\circ}$ angle, temperature $20\,^{\circ}{\rm C}$, ideal beam waist $16\,\mu$m). We can produce up to $1.6\, {\rm W}$ of blue light with a conversion efficiency larger than $50\%$~\cite{Delbino2015}. The blue light is locked to the atomic transition by performing saturated absorption spectroscopy in a see-through hollow-cathode lamp by Heraeus (buffer gas pressure $5\,{\rm torr}$). Although the residual laser linewidth is estimated to be well below $1\, {\rm MHz}$, the accuracy of the lock is of the order of a few MHz.
Light at $626\, {\rm nm}$ is generated by a commercial laser system (TA-SHG pro by Toptica, $1\, {\rm W}$ at $626\, {\rm nm}$, linewidth $20\, {\rm kHz}$ in $5 \,\mu$s). We lock the laser light to a Iodine transition using saturated absorption spectroscopy in a I$_2$ cell. The frequency gap between the locking point and the atomic transition (about $1\, {\rm GHz}$ for $^{162}$Dy) is bridged with a sequence of AOMs. The residual laser linewidth is a few tens of kHz. Fig.~\ref{Fig:lock} shows the relative position of the Dy  and Iodine absorption lines. This lock on a Iodine line represents a cost-effective alternative to the lock on an Ultra Low Expansion cavity. {Unfortunately, due to relative positions of Dysprosium and Iodine lines and to available AOMs, it gives access only to one of the two most abundant boson isotopes, i.e. $^{162}$Dy.} The laser light is conveyed to the different parts of the apparatus by single-mode, polarization-maintaining optical fibers.

\begin{figure}[htbp]
\centering{
\resizebox{0.75\columnwidth}{!}{
  \includegraphics{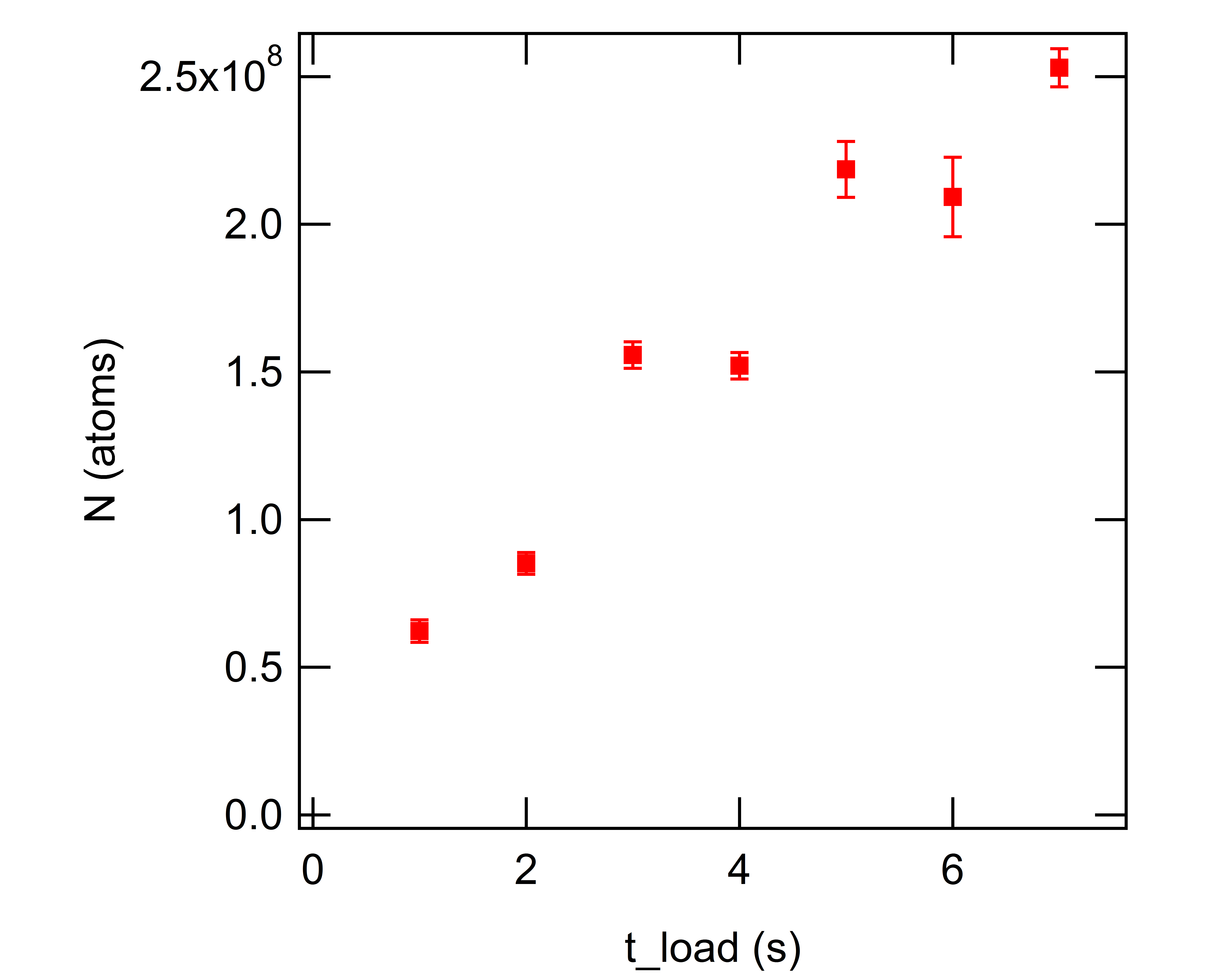}
	\includegraphics{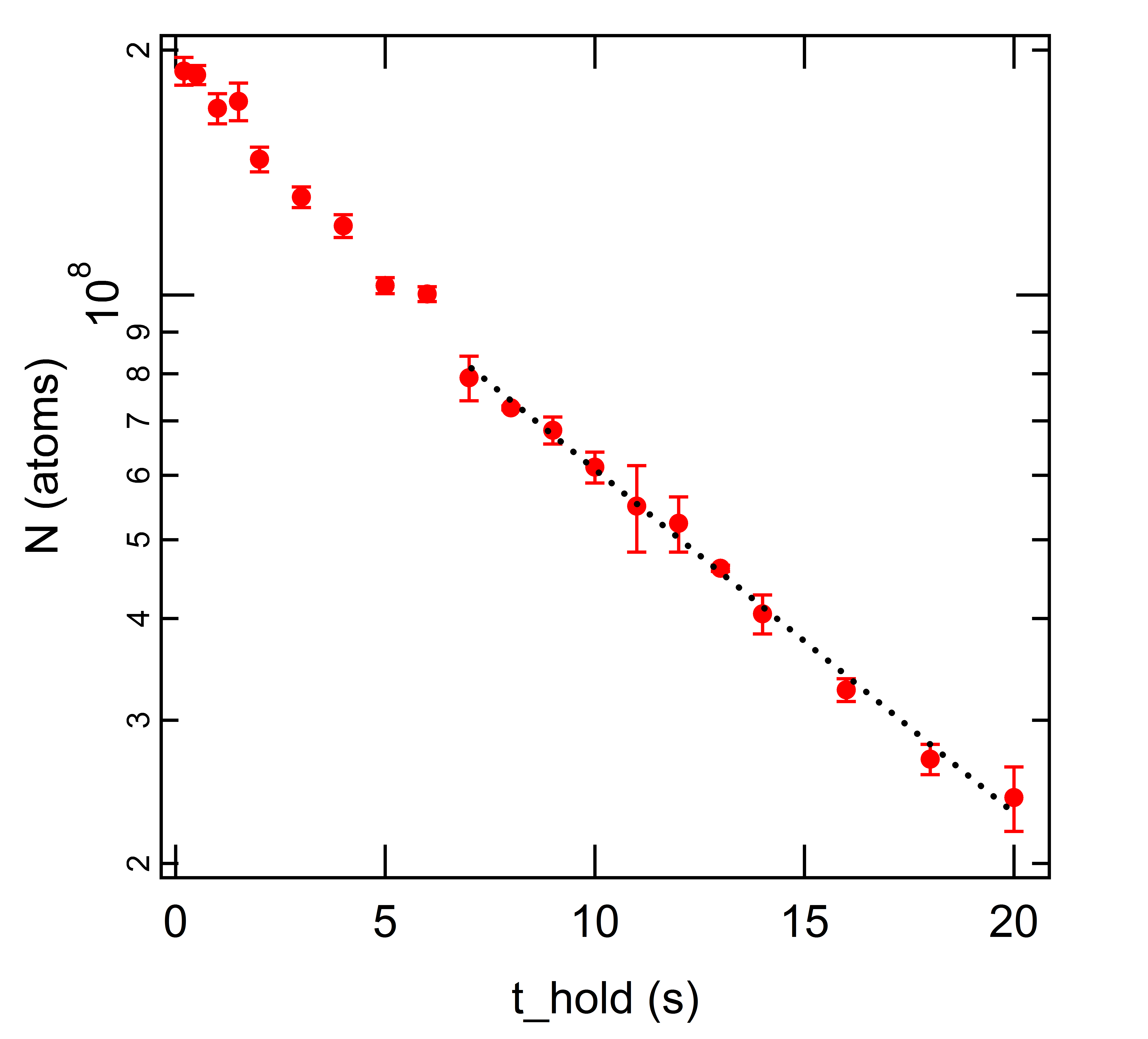}   }
}
\caption{(left part) Number of atoms in the MOT as a function of loading time; (right part) Number of remaining atoms in the MOT as function of the holding time; the line is a fit with a single exponential (see text). In both cases $\Delta=-35 \Gamma$ and total laser power on the ZS is $130\, {\rm mW}$.}
\label{fig:loading}       
\end{figure}

%
In the following,the measurements performed in order to optimize and characterize the MOT loading are reported.  Absorption imaging with resonant light at $421\, {\rm nm}$ along two orthogonal directions allows to extract the MOT dimensions and trapped atom number.  A compression phase lasting $40\, {\rm ms}$ at reduced  laser detuning ($\Delta_{\rm r} = -10\Gamma_{421}$) and intensity ($I_{\rm r} = 0.4 I$), is performed after the MOT phase, prior to imaging, in order to increase the density and decrease the temperature.

The narrow linewidth of the MOT transition allows, in our experimental conditions, a very low capture velocity of about $5\, {\rm m/s}$, as estimated from numerical simulations. In order to increase it, we have artificially increased the linewidth of the trapping laser during the loading phase by applying a frequency modulation at {a rate of} $136\, {\rm kHz}$ and amplitudes up to $2.0\, {\rm MHz}$, as performed in Ref.~\cite{Dreon2016}. This results in up to a factor 3 increase in the number of trapped atoms.
The number of atoms as a function of the loading time by the ZS and of the holding time in the MOT are shown in Fig.~\ref{fig:loading}. 
The atom capture rate at early times is $4.6\times 10^7\,{\rm s}^{-1}$ and the MOT atom number saturates after about $10\, {\rm s}$.
Typical densities, after the compression phase, are of the order of $10^{11}\,{\rm cm}^{-3}$.
We notice that the decay curve can be nicely fitted with a single exponential (decay time $\tau = 10.3\,{\rm s}$) only at long times, meaning that in our conditions the MOT density and the total loss rate are already affected by 2-body collisions. The temperature of the trapped atoms has been measured by absorption imaging after switching-off the MOT and letting the cloud expand freely for up to $30\, {\rm ms}$, obtaining temperatures of about $20\,\mu$K. 

\begin{figure}[htbp]
\centering{
\resizebox{0.75\columnwidth}{!}{
  \includegraphics{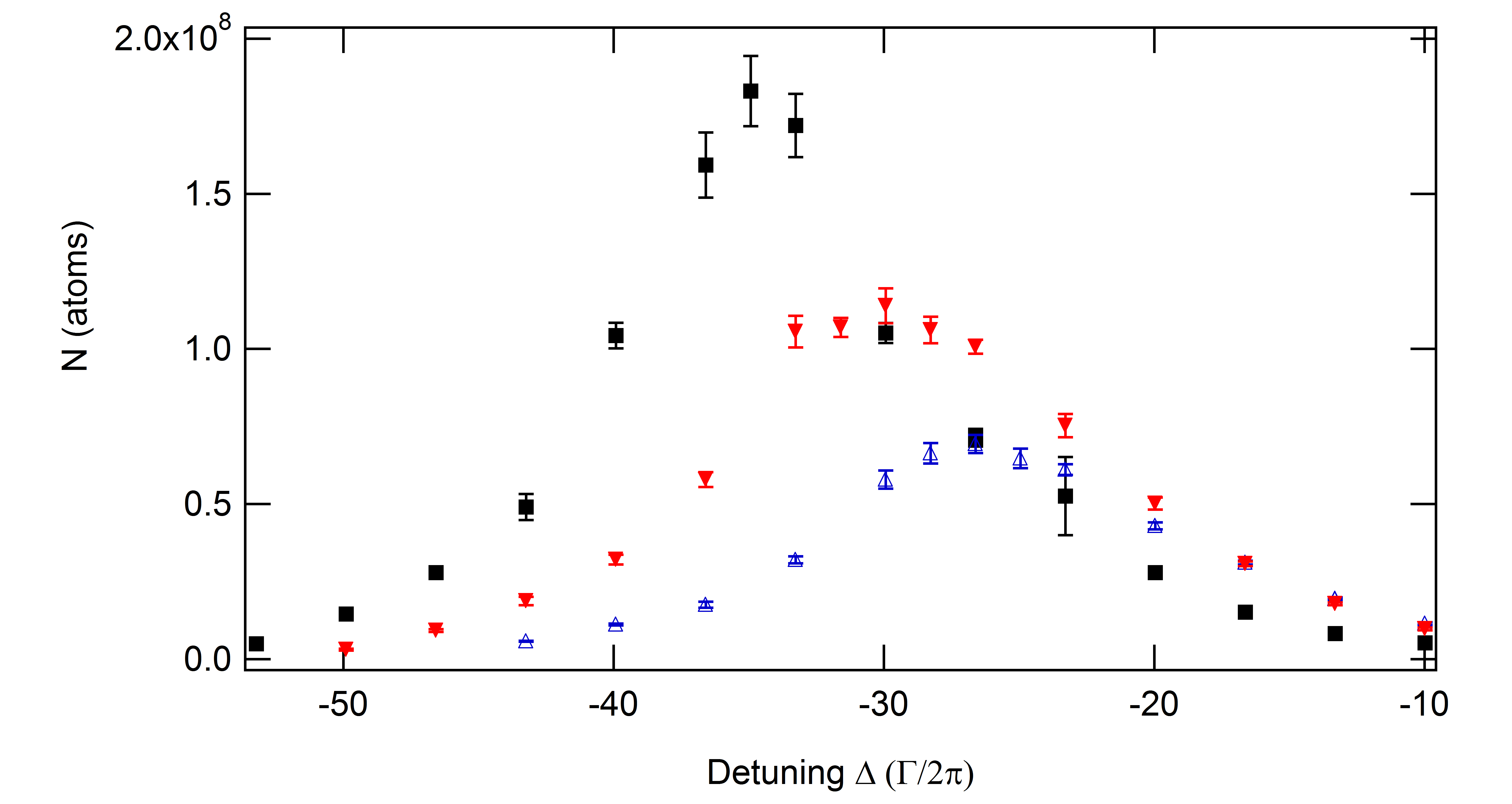} }
}
\caption{Number of atoms in the MOT as function of the detuning in the loading phase respectively with: (full squares)  $2.0\, {\rm MHz}$ laser linewidth, (full triangles) $1.1\, {\rm MHz}$ laser linewidth, (empty triangles) un-modulated laser, linewidth of a few tens kHz.}
\label{Fig:detuning}       
\end{figure}

A study of the number of trapped atoms as a function of MOT laser detuning and modulation band-width is shown in Fig.~\ref{Fig:detuning}. The observed trend is  consistent with that observed in similar setups~\cite{Frisch2012,Maier2014,Dreon2016}, i.e. optimum operating conditions at large detuning values. This feature can be ascribed to the detuning-dependent gravitational sag of the MOT, larger than $1\, {\rm cm}$ for large detunings, which ``protects'' the MOT against losses due to the ZS light during loading.  We observed that the trapped sample becomes more and more spin-polarized as the detuning is increased, due to the combined action of the gravitational sag and optical pumping. This is a particularly useful feature, studied in detail in Ref.~\cite{Dreon2016}, almost unique to this type of MOTs. We observed that spin-polarization can be reduced or lost during the compression phase if a small detunings or large intensities are employed.
As an example, the data in Fig.~\ref{fig:loading} are not completely polarized.

In conclusion, we are able to trap in our setup over 2$\times 10^{8}$ $^{162}$Dy atoms at temperatures as low as $20 \, \mu$K. This is a very good starting point for the subsequent evaporative cooling phase to quantum degeneracy. To this aim we have already included in our science cell a high-finesse in-vacuum Fabry-Perot cavity to {realize} a large-waist, standing-wave optical trap with a low power Nd-YAG laser. {A similar scheme has been already employed for Li and Yb atoms~\cite{Zimmermann,Pagano}.} This {method allows to realize a deep optical trap with a large volume, healing the problem of the relatively low dipole polarizability of Dy atoms at $1\,\mu$m wavelength~\cite{Dreon2016} and enhancing the capture of atoms from the MOT.} 
{After an initial evaporation stage, the atoms will be transferred to a standard crossed-beams optical trap and cooled evaporatively to quantum degeneracy. We plan to perform a first generation of experiments in the steel chamber, which provides sufficient optical access to realize optical lattices and other types of optical potentials (i.e. disorder or time-dependent potentials). For future experiments in well-controlled magnetic fields and with high-resolution imaging we plan to use a second glass chamber.}

{A first direction we plan to explore is that of low dimensional systems in ordered and disordered optical lattices. Working with short-wavelength lattices it is possible to enhance the effects of the dipolar interaction over the contact interaction~\cite{Baier2016}. In such regime, there are elusive and exotic quantum phases that could be addressed experimentally, such as supersolidity~\cite{Goral}, crystallization~\cite{Gopala} and many-body localization~\cite{Yao}}.

We thank M.~Tagliaferri, M.~Voliani, F.~Pardini, A.~Barbini, {R.~Ballerini, A.~Hajeb}, M.~De Pas, M.~Giuntini {and M.~Archimi} for technical assistance during the construction of the apparatus, {F.~Ferlaino, G.~Roati and D.~Giulietti for interesting discussions,} L.~Fallani for providing the ZS and MOT simulation programs, and L.A.~Gizzi for continuous support. {This work was supported also by the ERC (Grants 203479 - QUPOL and No. 247371 - DISQUA), by the EC - H2020 research and innovation programme (Grant No. 641122 - QUIC)}.

%
%
%
%

\begin{thebibliography}{}
\bibitem{Lahaye2009}
T.~Lahaye {\it et al.}, Rep. Prog. Phys. \textbf{72}, (2009) 126401
\bibitem{Trefzger2011}
{C.~Trefzger {\it et al.}, J. Phys. B \textbf{44}, (2011) 193001}
\bibitem{Baranov2012}
{M.~Baranov, Chem. Rev. \textbf{112}, (2012) 5012}
\bibitem{Lewenstein2007}
M.~Lewenstein {\it et al.}, Adv. Phys. \textbf{56}, (2007) 243
\bibitem{Bloch2008}
I.~Bloch {\it et al.}, Rev. Mod. Phys. \textbf{80}, (2008) 885
\bibitem{Lu2011a}
M.~Lu, {\it et al.}, Phys. Rev. Lett. \textbf{107}, (2011) 190401
\bibitem{Lu2012a}
M.~Lu, {\it et al.}, Phys. Rev. Lett. \textbf{108}, (2012) 215301
\bibitem{Aikawa2012}
K.~Aikawa, {\it et al.}, Phys. Rev. Lett. \textbf{108}, (2012) 210401
\bibitem{Aikawa2014}
K.~Aikawa, {\it et al.}, Phys. Rev. Lett. \textbf{112}, (2014) 010404
\bibitem{Baumann2014}
K.~Baumann, {\it et al.}, Phys. Rev. A \textbf{89}, (2014) 020701(R)
\bibitem{Maier2015a}
T.~Maier, {\it et al.}, Phys. Rev. A \textbf{92}, (2015) 060702(R)
\bibitem{Maier2015b}
T.~Maier, {\it et al.}, Phys. Rev. X \textbf{5}, (2015) 041029
\bibitem{Kadau2016}
H.~Kadau, {\it et al.}, Nature \textbf{530}, (2016) 194
\bibitem{Chomaz2016}
{L.~Chomaz, {\it et al.}, Phys. Rev. X \textbf{6}, (2016) 041039}
\bibitem{Baier2016}
S.~Baier, {\it et al.}, Science \textbf{352}, (2016) 201
\bibitem{Berglund2008}
A.J.~Berglund, {\it et al.}, Phys. Rev. Lett. \textbf{100}, (2008) 113002
\bibitem{Lu2010}
M.~Lu, {\it et al.}, Phys. Rev. Lett. \textbf{104}, (2010) 063001
\bibitem{Frisch2012}
A.~Frisch, {\it et al.}, Phys. Rev. A \textbf{85}, (2012) 051401(R)
\bibitem{Maier2014}
T.~Maier, {\it et al.}, Opt. Lett. \textbf{39}, (2014) 3138
\bibitem{Dreon2016}
D.~Dreon, {\it et al.}, arXiv:1610.02284v1 [cond-mat.quant-gas]
\bibitem{Delbino2015}
L.~Del~Bino, Master Thesis, Univ. Firenze, 2015, http://quantumgases.lens.unifi.it/publications/theses
\bibitem{Zimmermann} B. Zimmermann, {\it et al.}, New Jour. Phys. 13, (2011), 043007
\bibitem{Pagano} G. Pagano, {\it et al.}, Nat. Phys. 10, (2014) 198–201
\bibitem{Goral} K. Goral et al., Phys. Rev. Lett. 88, (2002), 170406
\bibitem{Gopala} S. Gopalakrishnan et al., Phys. Rev. Lett. 111, (2013), 185304
\bibitem{Yao} N. Y. Yao et al., Phys. Rev. Lett. 113, (2014), 243002
    
%
\end{thebibliography}
%

\end{document}